\documentclass[english]{article}
\usepackage[T1]{fontenc}
\usepackage[latin9]{inputenc}
\usepackage{geometry}
\usepackage{amsmath}
\usepackage{graphicx}
\usepackage{esint}
\usepackage[dvips]{color}
\usepackage[dvipsnames]{xcolor}
\def\nn{\nonumber}
\def\Q{\mathcal{Q}}

\newcommand{\al}{\alpha}

\newcommand{\pp}{{\textstyle \frac{\pi i} 2}}
\makeatletter

\providecommand{\tabularnewline}{\\}

\numberwithin{equation}{section}
\newcommand{\lyxaddress}[1]{
	\par {\raggedright #1
	\vspace{1.4em}
	\noindent\par}
}

\makeatother

\usepackage{babel}
\begin{document}
\title{Diagonal finite volume matrix elements in the sinh-Gordon model}
\author{Zoltan Bajnok$^{1}$, Fedor Smirnov$^{2}$\footnote{Membre du CNRS}}
\maketitle

\lyxaddress{\begin{center}
\emph{1. MTA Lend\"ulet Holographic QFT Group, Wigner Research Centre
for Physics,}\\
\emph{Konkoly-Thege Mikl\'os u. 29-33, 1121 Budapest, Hungary}\\
\emph{and}\\
\emph{2. Sorbonne Universite, UPMC Univ Paris 06 
CNRS, UMR 7589, LPTHE}\\ 
\emph{F-75005, Paris, France}
\par\end{center}}
\begin{abstract}
Using the fermionic basis we conjecture exact expressions for diagonal
finite volume matrix elements of exponential operators and their descendants
in the sinh-Gordon theory. Our expressions sum up the LeClair-Mussardo
type infinite series generalized by Pozsgay for excited state expectation
values. We checked our formulae against the Liouville three-point
functions for small, while against Pozsgay's expansion for large volumes
and found complete agreement. 
\end{abstract}

\section{Introduction}

Integrable models are ideal testing grounds of various methods and
ideas in quantum field theories. The simplest interacting model of
this type is the sinh-Gordon theory, which has a single particle type
and the full finite volume energy spectrum can be calculated from
the scattering phase of these particles. There is a hope that similar
exact results can be obtained also for finite volume matrix elements. 

The finite volume matrix elements of local operators are essentially
the building blocks of finite volume correlations functions, which
are relevant in statistical and solid state systems \cite{Mussardo:2010mgq,Samaj:2013yva}.
Their non-local counterparts can be used in the AdS/CFT correspondence
to describe three-point functions in the gauge theory and the string
field theory vertex in string theory \cite{Beisert:2010jr,Basso:2015zoa,Bajnok:2015hla,Basso:2017muf,Bajnok:2017mdf}.
Diagonal matrix elements play a special role there, as they describe
the HHL type correlation functions \cite{Bajnok:2014sza,Bajnok:2016xxu,Hollo:2015cda,Jiang:2015bvm,Jiang:2016dsr}. 

There were two alternative approaches for the calculation of finite
volume matrix elements. For generic operators and theories one can
try to use the infinite volume form factors \cite{Smirnov:1992vz,Babujian:2003sc}
and the scattering matrix \cite{Zamolodchikov:1978xm} to develop
a systematic large volume expansion. Polynomial volume corrections
originate from momentum quantization \cite{Luscher:1986pf}, while
exponentially small finite size corrections from the presence of virtual
particles \cite{Luscher:1985dn}. The LeClair-Mussardo formula \cite{Leclair:1999ys}
provides an infinite series for the exact finite volume one-point
function, where each term contains the contribution of a given number
of virtual particles in terms of their infinite volume connected form
factors and a weight function, which is related to the Thermodynamic
Bethe ansatz (TBA) densities of these particles \cite{Zamolodchikov:1989cf}.
This formula was then generalized by analytical continuation for diagonal
matrix elements, which replaces the ground-state TBA densities with
the excited state ones and contains additional factors, which can
be interpreted as partial density of states \cite{Pozsgay:2013jua,Pozsgay:2014gza}. 

Alternatively, there is an other approach which focuses on specific
theories and exploits their hidden (Grassmann) structure to provide
compact expressions for finite volume matrix elements \cite{HGSV}.
These specific continuum models arise as limits of integrable lattice
models and the most studied examples are the sinh-Gordon and sine-Gordon
models. There have been active work and relevant progress in deriving
finite volume one-point functions for the exponential operators and
their descendants in these theories \cite{Negro:2013rxa,Negro:2013wga}.
These results were then extended for diagonal matrix elements in the
sine-Gordon theory \cite{Hegedus:2017muz,Hegedus:2017zkz,Hegedus:2019zks}
and the aim of our paper is to provide similar expressions in the
sinh-Gordon theory. 

The paper is organized as follows: Section 2 reviews the description
of the finite size energy spectrum of the sinh-Gordon theory.  A multi-particle
state for large volumes can be labelled by momentum quantum numbers,
which we relate at small volume to the spectrum of the 
Liouville conformal field theory by matching the eigenvalues of the
conserved charges. In Section 3 we formulate our main conjecture for
the finite volume exceptions values in the fermionic basis. The novelty
compared to the vacuum expectation values is the discrete part of
the convolutions, which carries information on the particles' rapidities.
We check this conjecture for large volumes in Section 4. The discrete
part of the convolution contains the polynomial, while the continuous
part the exponentially small corrections in the volume. In Section
5 we compare our conjecture with Liouville three-point functions for
low lying states including non-degenerate and degenerate $L_{0}$
subspaces. All the checks performed confirm our conjecture, thus we
close the paper with conclusions in Section 6. 

\section{Energy spectrum}

In this section we summarize the exact description of the finite volume
energy spectrum together with its large and small volume formulations \cite{Teschner:2007ng}. 

The sinh-Gordon theory  is defined by the Lagrangian:
\begin{equation}
\mathcal{L}=\frac{1}{4\pi}(\partial\phi)^{2}+\frac{2\mu^{2}}{\sin\pi b^{2}}\cosh(b\varphi)\,.
\nn\end{equation}
In the literature there is an abundance of notations for the parameters of this model.
We decided to follow the paper \cite{Zamolodchikov:2000kt} by introducing
 the background charge of the related Liouville model and the renormalized coupling
constant as
$$Q=b+b^{-1}\,,\qquad p=\frac{b^2}{1+b^2}\,.$$
The sinh-Gordon model is the simplest integrable interacting two dimensional quantum field
theory. It has one single particle of mass $m$ with the corresponding
two particle scattering matrix 
\begin{equation}
S(\theta)=\frac{\sinh\theta-i\sin(\pi p)}{\sinh\theta+i\sin(\pi p)}\,\,.
\nn\end{equation}
The finite size energy spectrum, in a volume $R$, can be formulated
in terms of the $\Q$ function, which satisfies the following functional relations:
\begin{equation}
\Q(\theta+{\textstyle \frac{i\pi}{2}})\Q(\theta-{\textstyle \frac{i\pi}{2}})=1+\Q(\theta+{\textstyle \frac{i\pi}{2}}(1-2p))\Q(\theta-{\textstyle \frac{i\pi}{2}}(1-2p))\equiv1+e^{-\epsilon(\theta)}\,,\nn
\end{equation}
where we introduced the TBA pseudo-energy $\epsilon.$ Excited states
can be labeled by the zeros of $\Q$ as: $\{\theta_{1},\dots,\theta_{N}\}$.
With the prescribed large $\theta$ asymptotics, $\log \Q(\theta)\simeq-\frac{r}{2}\frac{\cosh\theta}{\sin\pi p}$
there is a unique solution
\begin{equation}
\Q(\theta)=\prod\limits _{k=1}^{N}\tanh\Bigl(\frac{\theta-\theta_{k}}{2}\Bigr)\exp\Bigl(-\frac{r\cosh(\theta)}{2\sin(\pi p)}+\frac{1}{2\pi}\int\limits _{-\infty}^{\infty}\frac{1}{\cosh(\theta-\theta')}\log\ensuremath{(1+e^{-\epsilon(\theta')})}d\theta'\Bigr)\,.
\nn\end{equation}
Here $r=mR$ is the dimensionless volume. Thanks to the functional
relation $\epsilon(\theta)$ can be fixed from the following TBA equation
\begin{align}
\epsilon(\theta)=r\cosh\theta+\sum\limits _{k=1}^{N}\log S(\theta-\theta_{k}-\pp
)-\int\limits _{-\infty}^{\infty}K(\theta-\theta')\log(\ensuremath{1+e^{-\epsilon(\theta')})}d\theta'\,.\label{TBA}
\end{align}
where the kernel is related to the scattering matrix as
\begin{equation}
K(\theta)=\frac{1}{2\pi i}\frac{d}{d\theta}\log S(\theta)=\frac{1}{2\pi i}\Bigl(\frac{1}{\sinh(\theta-\pi ip)}-\frac{1}{\sinh(\theta+\pi i p)}\Bigr)\,.
\nn\end{equation}
The finite size spectrum can be characterized by a set of integers
$\{N_{k}\}$, denoted by $\mathcal{N}$, via the zeros of the $\Q$
function, written equivalently as 
\begin{align}
f(\theta_{k})=\pi N_{k}\,,\label{BY}
\end{align}
 where 
\begin{equation}
f(\theta)=r\sinh\theta+\sum\limits _{k=1}^{N}\arg\ensuremath{(-S(\theta-\theta_{k}))}-\int\limits _{-\infty}^{\infty}K(\theta-\theta'+
\pp
)\log(\ensuremath{1+e^{-\epsilon(\theta')})}d\theta'\,.
\nn\end{equation}
coincides at the positions $\theta_{k}$ with the analytical continuation
of $-i\epsilon(\theta+i\pi/2)$ (with certain choice of the branches
of logarithms). 
Here we use $-S(\theta-\theta_{k})$ under $\arg$ for computational convenience (notice that $-S(0)=1$).

These equations are called the Bethe Ansatz (BA) equations
and can be interpreted as the momentum quantization equations of the
particles with rapidity $\theta_{k}$. The energy of the multi-particle
state with rapidities $\{\theta_{1},\dots,\theta_{N}\}$ can be written
as 
\begin{equation}
E_{\mathcal{N}}=\sum_{i=1}^{N}m\cosh\theta_{i}-m\int_{-\infty}^{\infty}\cosh\theta\log(1+e^{-\epsilon(\theta)})\frac{d\theta}{2\pi}\,.\label{eq:TBAenergy}
\end{equation}

\subsection{Large volume expansion}

Since in the large volume limit the TBA pseudo-energy behaves as $\epsilon=r\cosh\theta+O(1)$
the integral terms are of order $O(e^{-r})$
and can be neglected. This results in the large volume limit of the
BA equations 
\begin{equation}
r\sinh\theta_{j}+\sum\limits _{k=1}^{N}\arg\ensuremath{(-S(\theta_{j}-\theta_{k}))}=\pi N_{j}\,.\label{eq:BAlargeR}
\end{equation}
Let us assume that rapidities are labeled such that $\{\theta_{1}>\theta_{2}>\dots>\theta_{m}\}$.
We recall that in \cite{Teschner:2007ng} it was proven that for any
given set of integers $\{n_{1},\dots,n_{m}\}$ the equations
\begin{equation}
r\sinh\theta_{j}-\sum_{k=1}^{j-1}\arg(S(\theta_{k}-\theta_{j}))+\sum_{k=j+1}^{N}\arg(S(\theta_{j}-\theta_{k}))=2\pi n_{j}\,,\label{eq:BAargS}
\end{equation}
have a unique solution\footnote{Note that, although here the quantum numbers $\{n_{j}\}$ can be equal,
the solutions for the rapidities $\{\theta_{j}\}$ can not, so the
system is nevertheless fermionic and not bosonic type.}. The idea of the proof was to introduce $P(\theta)=\int_{0}^{\theta}\arg(S(v))dv$
and to show that the rapidities $\{\theta_{j}\}$ minimize the positive
definite Yang-Yang functional
\begin{equation}
\sum_{j}(r\cosh\theta_{j}-2\pi n_{j}\theta_{j})+\sum_{j<k}P(\theta_{j}-\theta_{k})\,.
\nn\end{equation}
In order to compare eq. (\ref{eq:BAlargeR}) to eq. (\ref{eq:BAargS})
we recall that for positive arguments 
\begin{equation}
\arg(S(\theta))=\arg(-S(\theta))-i\pi\quad;\qquad\theta>0\,.
\nn\end{equation}
This leads to the following relation between the quantum numbers $\{N_{j}\}$
and $\{n_{j}\}$: 
\begin{equation}
N_{j}=2n_{j}-N-1+2j\,.
\nn\end{equation}
In particular, the state labelled by $\{0,\dots,0\}$ in eq. (\ref{eq:BAargS})
will be mapped to $\{-M+1,-M+3,\dots,M-3,M-1\}$, with all quantum
numbers being distinct. This also shows that states with even number
of particles are labelled by odd quantum numbers, while states with
odd number of particles with even quantum numbers in $\mathcal{N}$.
Once the equations (\ref{eq:BAlargeR}) are solved for the rapidities
the large volume energy is 
\begin{equation}
E_{\mathcal{N}}=\sum_{i=1}^{N}m\cosh\theta_{i}\,.
\nn\end{equation}
In the following we analyze the small volume limit of the energies. 

\subsection{Small volume limit}

In the small volume limit we compare the energy eigenvalues with the
spectrum of the Liouville theory \cite{Zamolodchikov:1995aa}. In
this description the sinh-Gordon theory is understood as the perturbation
of the Liouville theory with the operator $e^{-b\varphi}$: 
\begin{equation}
\mathcal{L}=\mathcal{L}_{CFT}+\frac{\mu^{2}}{\sin\pi b^{2}}e^{-b\varphi}\,.
\nn\end{equation}
There are infinitely many conserved charges and the energy is related
to the first two as $E=-\frac{\pi}{12R}(I_{1}+\bar{I}_{1})$, where
\begin{equation}
I_{1}=-\frac{6r}{\pi}\biggl(\sum\limits _{k=1}^{N}e^{\theta_{k}}-\int\limits _{-\infty}^{\infty}e^{\theta}\log(\ensuremath{1+e^{-\epsilon(\theta)}})\frac{d\theta}{2\pi}\biggr)\,,\quad\bar{I}_{1}=-\frac{6r}{\pi}\biggl(\sum\limits _{k=1}^{N}e^{-\theta_{k}}-\int\limits _{-\infty}^{\infty}e^{-\theta}\log(\ensuremath{1+e^{-\epsilon(\theta)}})\frac{d\theta}{2\pi}\biggr)\,.\nn
\end{equation}
 The Liouville theory is a conformal field theory with a continuous
spectrum. Its Hilbert space is built up from the non-compact zero
mode and the oscillators. The zero mode determines the dimension of
the primary fields, while the oscillators create descendants. Once
the perturbation is introduced the spectrum of primary fields can
be approximated by the quantization of the zero mode \cite{Zamolodchikov:1995aa}:
\begin{align}
 & 4P_{L}(r)Q\log\Bigl(Z(p)rb^{\frac{b^{2}-1}{b^{2}+1}}\Bigr)=-\pi L+\frac{1}{i}\log\ensuremath{\frac{\Gamma(1+2iP_{L}(r)b)\Gamma(1+2iP_{L}(r)/b)}{\Gamma(1-2iP_{L}(r)b)\Gamma(1-2iP_{L}(r)/b)}}\,.\label{primary}
\end{align}
with $L=1,2,\dots$. Here and later we use the mass scale
 \begin{equation}
Z(p)=\frac{1}{16Q\pi^{3/2}}\Gamma\Bigl(\frac{p}{2}\Bigr)\Gamma\Bigl(\frac{1-p}{2}\Bigr).
\nn\end{equation}
 The eigenvalues of the conserved charges in the CFT can be written
as 
\begin{equation}
I_{1}^{\mathrm{CFT}}=P_{L}(r)^{2}-\frac{1}{24}+M\,,\quad\bar{I}_{1}^{\mathrm{CFT}}=P_{L}(r)^{2}-\frac{1}{24}+\overline{M}\,,
\nn\end{equation}
where $M,\overline{M}$ are levels of descendants for the two chiralities.
By comparing the energies in the TBA and the perturbed Liouville descriptions
we can relate the quantum numbers ${\cal N}$ to $\{L,M,\bar{M}\}$. 

To give an example, we claim that $\mathcal{N}=\{-2,0,2\}$ corresponds
to the primary field with $L=4$. Indeed, by numerically solving the
TBA and BA equations (\ref{TBA},\ref{BY}) on the one hand and the
zero mode quantization (\ref{primary}) on the other we found for
$r=.001$ the ratios: 
\begin{equation}
\frac{I_{1}}{I_{1}^{\mathrm{CFT}}}=1.00003\,,\quad\frac{\bar{I}_{1}}{\bar{I}_{1}^{\mathrm{CFT}}}=0.999989\,.
\nn\end{equation}
In this way we obtain the following correspondence, which we present
both at the language of $\mathcal{N}$ and that of $\{n_{j}\}$ with
$n_{j}=(N_{j}+M+1)/2-j$: 
\begin{align}
\begin{tabular}{|c|c|c|c|c|}
\hline  \ensuremath{\mathcal{N}}  &  \{\ensuremath{n_{i}}\}  &  \ensuremath{L}  &  \ensuremath{M}  &  \ensuremath{\bar{M}}\\
\hline\hline  \{\ \}  &  \{\}  &  1  &  0  &  0 \\
\hline  \{0\}  &  \{0\}  &  2  &  0  &  0 \\
\hline  \{-1,1\}  &  \{0,0\}  &  3  &  0  &  0 \\
\hline  \{-2,0,2\}  &  \{0,0,0\}  &  4  &  0  &  0 \\
\hline  \{2\}  &  \{1\}  &  1  &  1  &  0 \\
\hline  \{-1,3\}  &  \{0,1\}  &  2  &  1  &  0 \\
\hline  \{-2,0,4\}  &  \{0,0,1\}  &  3  &  1  &  0 \\
\hline  \{-3,3\}  &  \{-1,1\}  &  1  &  1  &  1 \\
\hline  \{4\}  &  \{2\}  &  1  &  2  &  0 \\
\hline  \{1,3\}  &  \{1,1\}  &  1  &  2  &  0 \\
\hline  \{-1,5\}  &  \{0,2\}  &  2  &  2  &  0 \\
\hline  \{-2,2,4\}  &  \{0,1,1\}  &  2  &  2  &  0 \\
\hline  \{-3,5\}  &  \{-1,2\}  &  1  &  2  &  1 \\
\hline  \{-5,5\}  &  \{-2,2\}  &  1  &  2  &  2 \\
\hline  \{1,5\}  &  \{1,2\}  &  1  &  3  &  0 \\
\hline  \{0,2,4\}  &  \{1,1,1\}  &  1  &  3  &  0 \\
\hline  \{-3,-1,1,5\}  &  \{0,0,0,1\}  &  4  &  1  &  0 \\
\hline 
\end{tabular}\label{tab}
\end{align}
Clearly in the parametrization $\{n_{j}\}$ the number of zeros is
$L-1$, while the sum of positive/negative numbers is $M/\bar{M}$,
in agreement with \cite{Teschner:2007ng}. 
Starting from $M=2$  the
spectrum of $L_0$ is degenerate. 
The 
degeneracy can be lifted using the second integral of motion as we demonstrate in Section \ref{degenerate}.
We will use all these states
later to compare our form factor conjecture to the Liouville three-point
functions in the small volume limit.

\section{Finite volume expectation values}

In the following we provide formulas for the expectation values
\begin{equation}
\langle\theta_{1
},\dots,\theta_{m}\vert\mathcal{O}\vert\theta_{m},\dots,\theta_{1}\rangle_{R}\,,
\nn\end{equation}
where $\vert\theta_{m},\dots,\theta_{1}\rangle_{R}$ is a normalized
finite volume energy eigenstate (\ref{eq:TBAenergy}) and $\mathcal{O}$
is a local operator. Expectation values of local operators obtained
by commuting with a conserved charge, $[I_{n},\mathcal{O}]$, vanish,
thus we consider only the quotient space, where these operators are
factored out. 

Local operators in massive perturbed conformal field theories are
in one-to-one correspondence with the states of the conformal Hilbert
space of the unperturbed model. The sinh-Gordon theory can be considered
either as the perturbation of the free massless boson with $\cosh(b\varphi)$
or as the perturbation of the Liouville theory with the operator $e^{-b\varphi}$.
Local operators are the exponentials $\Phi_{\alpha}=e^{\frac{Q\alpha}{2}\varphi}(0)$
together with their descendants $\mathcal{O}_{\alpha}$, which can
be generated in two different ways \cite{Fateev:1998xb,Negro:2013rxa,Negro:2013wga}.
If the modes of the free massless boson are used the operators are
called \emph{Heisenberg descendants} and the expectation values of
the corresponding operators in the quotient space have the $\sigma_{1}:\alpha\to-\alpha$
symmetry. In the perturbed Liouville scheme \emph{Virasoro descendants}
are generated by the Virasoro modes and the expectation values have
the symmetry $\sigma_{2}:\alpha\to2-\alpha$. 
Relating these two descriptions should provide a basis of the CFT adapted to
the integrable perturbation. Direct attempt to find such a basis failed for level higher than $2$
because it requires solving a rather complicated Riemann-Hilbert problem. The solution came from a
rather distant study of lattice integrable models which lead to the discovery of the fermionic basis. 
The latter provides in the scaling limit the fermionic basis for the sine-Gordon model
\cite{HGSIV,HGSV}. As has been shown
in \cite{Negro:2013rxa} this fermionic basis 
brings the Riemann-Hilbert problem in question to the diagonal form.

\subsection{Fermionic basis}

The definition of the fermionic basis in the CFT case can be considered as a purely algebraic one,
that is why it is equally suitable for the sinh-Gordon case. Analytical advantage of using the fermionic 
basis in the sine-Gordon model is due to the fact that the expectation values of the elements of the fermionic
basis are expressed as determinants. We do not know how to prove similar fact for the sinh-Gordon case, 
so, like in \cite{Negro:2013wga} we shall formulate it as a conjecture and then perform numerous checks. 

The  fermionic  basis is
created by the anti-commutative operators $\beta^{*},\gamma^{*}$
(and $\bar{\beta}^{*},\bar{\gamma}^{*}$ for the other chirality). They
can be used to generate the quotient space as 
\begin{equation}
\beta_{M}^{*}\gamma_{N}^{*}\bar{\beta}_{\bar{M}}^{*}\bar{\gamma}_{\bar{N}}^{*}\Phi_{\alpha}=\beta_{m_{1}}^{*}\dots\beta_{m_{k}}^{*}\gamma_{n_{1}}^{*}\dots\gamma_{n_{k}}^{*}\bar{\beta}_{\bar{m}_{1}}^{*}\dots\bar{\beta}_{\bar{m}_{\bar{k}}}^{*}\bar{\gamma}_{\bar{n}_{1}}^{*}\dots\bar{\gamma}_{\bar{n}_{\bar{k}}}^{*}\Phi_{\alpha}\,,\nn
\end{equation}
with all modes being odd and positive. 
Later we shall have these  operators for negative
indices. By definition they are related to annihilation operators $\beta_{-j}^{*}=\gamma_{j}$
, $\gamma_{-j}^{*}=\beta_{j}$ (together with similar relations for
the other chirality) such that their anti-commutator is 
\begin{equation}
\{\beta_{m},\beta_{n}^{*}\}=\{\bar{\gamma}_{m},\bar{\gamma}_{n}^{*}\}=-t_{m}(\alpha)\delta_{m,n}\quad;\qquad t_{n}(\alpha)=\frac{1}{2\sin{\pi}(\al-np)}\,.\nn
\end{equation}
The relation between the fermionic basis and the Heisenberg or Virasoro
basis is a very complicated problem and requires a case by case study. 
Later we shall have examples.

What is particularly nice about the fermionic basis is that the finite
volume expectation values take a very simple determinant form. Indeed,
the main result of our paper is a conjecture of the form 
\begin{equation}
\frac{\langle\theta_{1},\dots,\theta_{m}\vert\beta_{M}^{*}\gamma_{N}^{*}\bar{\beta}_{\bar{M}}^{*}\bar{\gamma}_{\bar{N}}^{*}\Phi_{\alpha}\vert\theta_{m},\dots,\theta_{1}\rangle_{R}}{\langle\theta_{1},\dots,\theta_{m}\vert\Phi_{\alpha}\vert\theta_{m},\dots,\theta_{1}\rangle_{R}}=\mathcal{D}(\{M\cup(-\bar{M})\}\vert\{N\cup(-\bar{N})\}\vert\alpha)\nn\,.
\end{equation}
where for the index sets $A=\{a_{1},\dots,a_{n}\}$ and $B=\{b_{1},\dots,b_{n}\}$
the determinant is
\begin{equation}
\mathcal{D}(A\vert B\vert\alpha)=\prod_{j=1}\frac{\text{sgn}(a_{j})\text{sgn}(b_{j})}{\pi}\text{Det}_{j,k}\left(\Omega_{a_{j},b_{k}}\right)\quad;\quad\Omega_{n,m}=\omega_{n,m}-\pi\text{sgn}(n)\delta_{n,-m}t_{n}(\alpha)\,.\nn
\end{equation}
The construction of the matrix $||\omega_{m,n}||$ is explained in the next section.

\subsection{The matrix $\omega_{m,n}$}
The matrix $\omega_{m,n}$ is built via a deformation of a linear operator
involved in the linearization of the TBA equations. We start by explaining this linearization. 
Consider the variation of  the TBA equations \eqref{TBA},\eqref{BY} with respect to $r$. 
The functions  $\epsilon(\theta)$ and $f(\theta)$ depend actually on $\theta$ and $r$, while the points of the discrete spectrum
$\theta_k$ depends on $r$. We have
\begin{align}
&\partial_r\epsilon(\theta)=\cosh\theta-2\pi i\sum\limits _{k=1}^{N}K(\theta-\theta_{k}+
\pp
)\frac{d \theta_k}{dr}+\int\limits _{-\infty}^{\infty}K(\theta-\theta')\partial_r\epsilon(\theta')\frac 1 {1+e^{\epsilon(\theta')}}d\theta'\label{variation}\\
&=\cosh\theta+2\pi i\sum\limits _{k=1}^{N}K(\theta-\theta_{k}+
\pp
)
\frac{1}{\partial_{\theta}f(\theta_k)}\partial_{r}f(\theta_k)+\int\limits _{-\infty}^{\infty}K(\theta-\theta')\partial_r\epsilon(\theta')\frac 1 {1+e^{\epsilon(\theta')}}d\theta'\,,\nn
\end{align}
where we used that $$\partial_{r}f(\theta_k)+\partial_{\theta}f(\theta_k)\frac{d \theta_k}{dr}=0\,,$$
following from \eqref{BY}.

Consider functions on discrete and continuous spectra $G=\{g_1,\cdots, g_k,g(\theta)\}$.
Motivated by \eqref{variation} we introduce paring for two such functions
\begin{equation}
G* H=2\pi i \sum\frac{1}{\partial_{\theta}f(\theta_k)}
g_k h_k+\int\limits_{-\infty}^{\infty}g(\theta)h(\theta)\frac{d\theta}{1+e^{\epsilon(\theta)}}\,.
\label{eq:convolution}
\end{equation}

By using this convolution the matrix element $\omega_{n,m}$ entering in our conjecture can be written as  
\begin{align}
\omega_{n,m} & =e_{n}*(1+\mathcal{K}_{\alpha}+\mathcal{K}_{\alpha}*\mathcal{K}_{\alpha}+\dots)*e_{m}\equiv e_{n}*(1+\mathcal{R}_{\mathrm{dress},\alpha})*e_{m}\nn\,,
\end{align}
where $e_{n}=\{e^{n(\theta_{1}+
\pp
)},\dots,e^{n(\theta_{m}+\pp
)},e^{n\theta}\}$
and $\mathcal{K}_{\alpha}$ has a matrix structure
\begin{equation}
\mathcal{K}_{\alpha}=\begin{pmatrix}K_{\alpha}(\theta_{k}-\theta_{l}) & K_{\alpha}(\theta_{k}-\theta+
\pp
)\\
K_{\alpha}(\theta-\theta_{l}-
\pp
) & K_{\alpha}(\theta-\theta')
\end{pmatrix}\nn\,.
\end{equation}
reflecting the fact that the convolution has a discrete and the continuous
part. Here $K_{\alpha}$ is the deformation of the TBA kernel 
\begin{equation}
K_{\alpha}(\theta)=\frac 1 {2\pi i}\Bigl(\frac{e^{-i\pi\alpha}}{\sinh(\theta-\pi ip)}-\frac{e^{i\pi\alpha}}{\sinh(\theta+\pi i p)}\Bigr)\,,\nn
\end{equation}
which satisfy $K_{\alpha+2}(\theta)=K_{\alpha}(\theta)$. 

Similar determinant expression to ours was proposed and tested for vacuum expectation values
in \cite{Negro:2013wga}. Our formulae are the extensions of VEVs
for excited states and the novel complication is the discrete part
of the convolutions. In the next section we explain how to work with
these expressions. 

There was a nice observation in \cite{HGSV} that one might
relax the condition that the number of $\beta^{*}$ and $\gamma^{*}$
are the same, but in the same time maintain the determinant form.
By this way operators with different sectors can be connected as 
\begin{equation}
\beta_{M}^{*}\gamma_{N}^{*}\bar{\beta}_{\bar{M}}^{*}\bar{\gamma}_{\bar{N}}^{*}\Phi_{\alpha+2mp}=\frac{C_{m}(\alpha)}{\prod_{j=1}^{m}t_{2j-1}(\alpha)}\beta_{M+2m}^{*}\gamma_{N-2m}^{*}\bar{\beta}_{\bar{M}-2m}^{*}\bar{\gamma}_{\bar{N}+2m}^{*}\beta_{\{m\}}^{*}\bar{\gamma}_{\{m\}}^{*}\Phi_{\alpha}\nn\,.
\end{equation}
 where $\{m\}=1,3,\dots,2m-1$ and $C_{m}(\alpha)$ is the ratio of
the infinite volume vacuum expectation values \cite{Fateev:1998xb}:
\begin{equation}
C_{m}(\alpha)=\frac{\langle\Phi_{\alpha-2mp}\rangle_{\infty}}{\langle\Phi_{\alpha}\rangle_{\infty}}\,.\nn
\end{equation}
The simplest of these relations is 
\begin{equation}
\frac{\Phi_{\alpha-2p}}{\langle\Phi_{\alpha-2p}\rangle_{\infty}}=\frac{1}{t_{1}(\alpha)}\beta_{1}^{*}\bar{\gamma}_{1}^{*}\frac{\Phi_{\alpha}}{\langle\Phi_{\alpha}\rangle_{\infty}}\,.\nn
\end{equation}
This relation is understood in the weak sense, i.e. for matrix elements.
In the next section we take diagonal matrix elements of this relation
and compare its large volume expansion with Pozsgay's result \cite{Pozsgay:2013jua}. 

\section{Large volume checks}

In this section we make some IR checks of our formulae for the diagonal
finite volume matrix elements, which we normalize as
\begin{equation}
F(\theta_{1},\dots,\theta_{m}\vert\alpha)=\frac{\langle\theta_{1},\dots,\theta_{m}\vert\Phi_{\alpha}\vert\theta_{m},\dots,\theta_{1}\rangle_{R}}{\langle\Phi_{\alpha}\rangle_{\infty}}\,.\nn
\end{equation}
Reflection properties with $\sigma_{1}$ and $\sigma_{2}$ ensure the invariance under the $\alpha\to\alpha+2$
shift. The finite volume state $\vert\theta_{m},\dots,\theta_{1}\rangle_{R}$
is symmetric in the rapidity variables, which satisfy the BA
equations $f(\theta_{k})=\pi N_{k}$. These states can be labelled either
by the discrete quantum numbers $N_{k}$ or by the rapidities
$\theta_{k}$ and are naturally normalized to Kronecker delta
functions. 
In the following we investigate the simplest non-trivial example
\begin{equation}
\frac{F(\theta_{1},\dots,\theta_{m}\vert\al-2p)}{F(\theta_{1},\dots,\theta_{m}\vert\alpha)}=1+\frac{2}{\pi}\sin\pi(p-\al)(e_1*e_{-1}+e_1*\mathcal{R}_{\mathrm{dress},\alpha}*e_{-1})\,.\label{eq:e1Rem1}
\end{equation}
where $e_n$ is related to $e^{n\theta}$. For each function $g$ we have a discrete and a continues part: $(g_{1},\dots,g_{m},g(\theta))$
with $g_{j}=g(\theta_{j}+i\frac{\pi}{2})$ and the convolution is
understood as in \eqref{eq:convolution}. We would like to compare these
formulae with the available results in literature which we recall
now.

\subsection{Form factor expansion of the diagonal finite volume matrix elements}

A \emph{finite} volume diagonal form factor can be expressed in terms
of the \emph{infinite }volume connected form factors, which are defined
to be the finite ($\epsilon$-independent) part in the crossed expression
\begin{equation}
F_{2n}^{\mathcal{O}}(\theta_{1}+i\pi+\epsilon_{1},\dots,\theta_{m}+i\pi+\epsilon_{m},\theta_{m},\dots,\theta_{1})=\frac{O(\epsilon^{m})}{\epsilon_{1}\dots\epsilon_{m}}+F_{2n,c}^{\mathcal{O}}(\theta_{1},\dots,\theta_{m})+O(\epsilon)
\,.\nn
\end{equation}
These connected form factors have interesting properties \cite{Pozsgay:2014gza}.
For the exponential operators, normalized by the VEV's, $\Phi_{\alpha}/\langle\Phi_{\alpha}\rangle_{\infty}$,
the first two connected form factors read as \cite{Leclair:1999ys,Negro:2013wga}:
\begin{align}
&F_{2,c}^{\alpha}=4\sin(\pi p)[k_{\alpha}]^{2}\,,\nn\\
&F_{4,c}^{\alpha}(\theta_1,\theta_2)=4\pi F_{2,c}^{\alpha}K(\theta_1-\theta_2)\Bigl(\cosh(\theta_1-\theta_2)[k_{\alpha}]^{2}-\frac{[k_{\alpha}-1][k_{\alpha}+1]}{\cosh(\theta_1-\theta_2)}\Bigr)\,,\nn
\end{align}
where
\[
[k]=\frac{\sin({\pi} p k)}{\sin({\pi}p)}\quad;\qquad k_{\alpha}=\frac{\alpha}{2p}\,.
\]
The six particle connected form factor
would fill a half page and there is no closed form available for the
general case. In principle such expressions could be obtained from
the determinant representation of form factors \cite{Koubek:1993ke}
using the limiting behavior of the symmetric polynomials \cite{Pozsgay:2014gza}.
But this procedure is quite cumbersome and the results do not seem to have any nice structure, thus  we restrict our investigations for these first two form factors only. 

The exact formula for the finite volume diagonal matrix elements was
conjectured in \cite{Pozsgay:2013jua} based on carefully evaluating
the contour deformation trick in the LM formula for 1 and 2 particle
states. For $m$ particles the conjecture takes the form 
\begin{equation}
F(\theta_{1},\dots,\theta_{m}\vert\alpha)=\frac{\sum_{I\subseteq M}\mathcal{F}_{m-\vert I\vert}^{\alpha}(M\setminus I)\rho_{\vert I\vert}(I)}{\rho_{m}(M)}\,,\label{eq:FVexact}
\end{equation}
where the full index set is denoted by $M=\{1,2,\dots,m\}$, and an
index set $I=\{i_{1},\dots,i_{k}\}$ in the argument abbreviates the
set 
of rapidities $\theta_{i_{1}},\dots,\theta_{i_{k}}$. The appearing
densities $\rho_{\vert I\vert}(I)$ are defined to be the determinant
\begin{equation}
\rho_{k}(\theta_{i_{1}},\dots,\theta_{i_{k}})=\det_{j,l}\left|\partial_{\theta_{i_{j}}}f(\theta_{i_{l}})\right|\,.\nn
\end{equation}
For $I=M$ this is simply the density of the finite volume $m$-particle
states. The quantity $\mathcal{F}_{k}^{\alpha}$ is the generalization of
the LM expansion for the connected form factor $F_{2k,c}^{\alpha}$:
\begin{equation}
\mathcal{F}_{k}^{\alpha}(\theta_{1},\dots,\theta_{k})=\sum_{n=0}^{\infty}\frac{1}{n!}\prod_{j=1}^{n}\int\frac{dm(v_{j})}{2\pi}F_{2(k+n),c}^{\alpha}(\theta_{1}+
\pp
,\dots,\theta_{k}+
\pp
,v_{1},\dots,v_{n})\,,\nn
\end{equation}
where $dm(v)=\frac{dv}{1+e^{\epsilon(v)}}$. 

\subsection{Checks at polynomial order}

The finite volume diagonal form factor at any polynomial order in
the inverse of the volume can be obtained from (\ref{eq:FVexact})
by neglecting the integral terms both in $f$ and also in $\mathcal{F}$.
At this order 
\begin{equation}
\mathcal{F}_{k}^{\alpha}(\theta_{1},\dots,\theta_{k})=F_{2k,c}^{\alpha}(\theta_{1},\dots,\theta_{k})+O(e^{-r})\,,\nn
\end{equation}
and 
\begin{equation}
\rho_{k}(\theta_{i_{1}},\dots,\theta_{i_{k}})=\det_{j,l}\left|\delta_{j,l}(r\cosh\theta_{j}+2\pi\sum_{n=1}^{m}K(\theta_{i_{j}}-\theta_{n}))-2\pi K(\theta_{i_{j}}-\theta_{i_{l}})\right|+O(e^{-r})\,,\nn
\end{equation}
This asymptotic expression was conjectured in \cite{Pozsgay:2007gx} based on form
factor perturbation theory and later proved in \cite{Bajnok:2017bfg}, moreover, it leads 
to the proof of the LM series, whose analytic continuation provided the exact conjecture (\ref{eq:FVexact}).
In the following we recover these asymptotic results from our fermionic expression (\ref{eq:e1Rem1}).
We proceed in the particle number. 

\subsubsection{1-particle case}

We need to check the relation
\begin{align}
\frac{F(\theta_{1}\vert\alpha -2p)}{F(\theta_{1}\vert\alpha)} & =\frac{F_{2,c}^{\alpha -2p}+r\cosh\theta_{1}}{F_{2,c}^{\alpha}+r\cosh\theta_{1}}+O(e^{-r})=\label{eq:1pff}\\
 & =1+\frac{2}{\pi}\sin\pi(p-\al)(e_1*e_{-1}+e_1*\mathcal{R}_{\mathrm{dress},\alpha}*e_{-1})\,,\nonumber 
\end{align}
where $e_1$ represents one discrete particle with $\theta_{1}$
and the function $e^{v_{1}}$ as $(e^{\theta_{1}+\frac{i\pi}{2}},e^{v_{1}})$.
Thus the first convolution in \eqref{eq:convolution} explicitly reads
as
\begin{equation}
e_1*e_{-1}=\frac{2\pi}{f'(\theta_{1})}+\int dm(v_{1})=\frac{2\pi}{r\cosh\theta_{1}+2\pi K(0)}+O(e^{-r})\,,\label{eq:1pLO}
\end{equation}
where we evaluated it neglecting exponentially small terms. Clearly,
each convolution in the discrete part introduces a polynomial suppression
factor $r^{-1}$ while in the continuous part an exponential one $e^{-r}$.
In order to obtain all polynomial volume corrections we need to sum
up the iterated series in the discrete part
\begin{align}
e_1*e_{-1}+e_1*\mathcal{R}_{\mathrm{dress},\alpha}*e_{-1} & =e_1*(1+K_{\alpha}(0)+K_{\alpha}(0)*K_{\alpha}(0)+\dots)*e_{-1}\label{eq:1pferm}\\
 & =\frac{2\pi}{f'(\theta_{1})}\left(1+\sum_{n=1}^{\infty}\frac{(2\pi K_{\alpha}(0))^{n}}{f'(\theta_{1})^{n}}\right)=\frac{2\pi}{f'(\theta_{1})-2\pi K_{\alpha}(0)}\nonumber \\
 & =\frac{2\pi}{r\cosh\theta_{1}+2\pi(K(0)-K_{\alpha}(0))}+O(e^{-r})\,.\nonumber 
\end{align}
The relation $2\pi(K(0)-K_{\alpha}(0))=F_{2,c}^{\alpha}$ together with $F_{2,c}^{\al-2p}-F_{2,c}^{\alpha}=4\sin(\pi(p-\al))$
imply that the two forms (\ref{eq:1pff}) and (\ref{eq:1pferm})
are indeed equivalent. 

\subsubsection{2-particle case}

For two particles the form factor expression, neglecting exponential
corrections, gives 
\begin{equation}
\frac{F(\theta_{1},\theta_{2}\vert\al-2p)}{F(\theta_{1},\theta_{2}\vert\alpha)}=\frac{F_{4,c}^{\al-2p}(\theta_{1},\theta_{2})+F_{2,c}^{\al-2p}(\rho_{1}(\theta_{1})+\rho_{1}(\theta_{2}))+\rho_{2}(\theta_{1},\theta_{2})}{F_{4,c}^{\alpha}(\theta_{1},\theta_{2})+F_{2,c}^{\alpha}(\rho_{1}(\theta_{1})+\rho_{1}(\theta_{2}))+\rho_{2}(\theta_{1},\theta_{2})}\,.\label{eq:2pff}
\end{equation}
In the fermionic formulation we consider only the discrete part, thus
in the convolution $e_1$ represents $i(e^{\theta_{1}},e^{\theta_{2}})$,
while $e_{-1}$ is nothing but $-i(e^{-\theta_{1}},e^{-\theta_{2}})$.
The kernel is a $2\times 2$ matrix: $(\hat{K}_{\alpha})_{ij}\equiv K_{\alpha}(\theta_{i}-\theta_{j})$.
The convolution in the discrete part can be traded for ordinary matrix
multiplication by introducing an extra matrix factor\footnote{Note that $f'(\theta_{j})-\partial_{\theta_{j}}f(\theta_{j})=2\pi K(0)+O(e^{-r})$.} $(\hat{f})_{ij}=\delta_{ij}f'(\theta_{j})$.
As a result we obtain 
\begin{equation}
\frac{F(\theta_{1},\theta_{2}\vert\al-2p)}{F(\theta_{1},\theta_{2}\vert\alpha)}=1+4\sin\pi(p-\al)(e^{\theta_{1}},e^{\theta_{2}})(\hat{f}-2\pi\hat{K}_{\alpha})^{-1}\text{\ensuremath{\left(\begin{array}{c}
 e^{-\theta_{1}}\\
 e^{-\theta_{2}} 
\end{array}\right)}}\,,\label{eq:2pferm}
\end{equation}
where exponential corrections are neglected, but all polynomial corrections
are summed up. We have checked explicitly that this result (\ref{eq:2pferm})
agrees with the form factor description (\ref{eq:2pff}). 

\subsubsection{m-particle case}

Similar calculation can be repeated for the generic $m$-particle
case. Now $\hat{K}_{\alpha}$ and $\hat{f}$ are $m\times m$ matrices
with entries 
\begin{equation}
(\hat{K}_{\alpha})_{ij}=K_{\alpha}(\theta_{i}-\theta_{j})\quad;\qquad(\hat{f})_{ij}=\delta_{ij}(r\cosh\theta_{j}+2\pi\sum_{k=1}^{m}K(\theta_{j}-\theta_{k}))\,,\nn
\end{equation}
leading to the analogous formula
\begin{align}
\frac{F(\theta_{1},\dots,\theta_{m}\vert\al-2p)}{F(\theta_{1},\dots,\theta_{m}\vert\alpha)} & =1+4\sin\pi(p-\al)(e^{\theta_{1}},\dots,e^{\theta_{m}})(\hat{f}-2\pi\hat{K}_{\alpha})^{-1}\text{\ensuremath{\left(\begin{array}{c}
 e^{-\theta_{1}}\\
 \vdots\\
 e^{-\theta_{m}} 
\end{array}\right)}}\,.\nn
\end{align}
Since higher than two-particle connected form factors are very complicated
we did not check explicitly this result, although we have no doubts
about its correctness. However, we would like to point out that substituting
$r=0$ in the formula provides a very compact and simple expression
for the ratios of diagonal matrix elements. These are actually nothing
but the symmetric evaluations of the form factors \cite{Pozsgay:2007gx}.
We believe that this observation could be used to find some nice parametrization
of these form factors in the generic case. 

\subsection{Checks at the leading exponential order }

We now check the leading exponential correction for the simplest 1-particle form factor
\begin{equation}
\frac{F(\theta_{1}\vert\al-2p)}{F(\theta_{1}\vert\alpha)}=\frac{F_{2,c}^{\al-2p}+\rho_{1}(\theta_{1})+\int\frac{dm(v_{1})}{2\pi}(F_{4,c}^{\al-2p}(\theta_{1}+\frac{i\pi}{2},v_{1})+\rho_{1}(\theta_{1})F_{2,c}^{\al-2p})}{F_{2,c}^{\alpha}+\rho_{1}(\theta_{1})+\int\frac{dm(v_{1})}{2\pi}(F_{4,c}^{\alpha}(\theta_{1}+\frac{i\pi}{2},v_{1})+\rho_{1}(\theta_{1})F_{2,c}^{\alpha})}+O(e^{-2r})\,,\label{eq:1pNLO}
\end{equation}
where we also need to expand $\rho_{1}(\theta_{1})$. In doing so
we recall that 
\begin{equation}
\rho_{1}(\theta_{1})=\partial_{\theta_{1}}f(\theta_{1})=r\cosh\theta_{1}-i\int dm(\theta)K(\theta_{1}+i\frac{\pi}{2}-\theta)\biggl(\frac{\partial\epsilon(\theta)}{\partial\theta}+\frac{\partial\epsilon(\theta)}{\partial\theta_{1}}\biggr)\,.\nn
\end{equation}
By differentiating the TBA equation wrt. to both $\theta_{1}$ and
$\theta$ we obtain linear integral equations with solutions

\begin{align}
&\frac{\partial\epsilon(\theta)}{\partial\theta}=r\sinh\theta+2\pi iK(\theta-\frac{i\pi}{2}-\theta_{1})+\int dm(v)R_{\mathrm{dress}}(\theta-v)(r\sinh v+2\pi iK(v-\frac{i\pi}{2}-\theta_{1}))\,,\label{eq:depstheta}
\\&
\frac{\partial\epsilon(\theta)}{\partial\theta_{1}}=-2\pi iK(\theta-\frac{i\pi}{2}-\theta_{1})-2\pi i\int dm(v)R_{\mathrm{dress}}(\theta-v)K(v-\frac{i\pi}{2}-\theta_{1}))\,,\label{eq:depstheta1}
\end{align}
 where the resolvent $R_{\mathrm{dress}}$ satisfies the equation
\begin{equation}
R_{\mathrm{dress}}(\theta)-\int dm(v)R_{\mathrm{dress}}(\theta-v)K(v)=K(\theta)\,.\nn
\end{equation}
 Thus at the leading exponential order 
\begin{equation}
\rho_{1}(\theta_{1})=r\Bigl(\cosh\theta_{1}-i\int dm(\theta)K(\theta_{1}+i\frac{\pi}{2}-\theta)\sinh\theta\Bigr)+O(e^{-2r})\,.\nn
\end{equation}
This allows us to expand the denominator and keep only the leading
exponential piece in order to compare with the formula coming from
the fermionic description (\ref{eq:e1Rem1}).

Evaluating the leading piece in the fermionic formula provides (\ref{eq:1pLO}).
To get the remaining terms we sum up the iterative terms. Keeping in mind
that $e_{1}$ represents the discrete and the continuous parts $(ie^{\theta_{1}},e^{v_{1}})$
the $k^{th}$ convolution gives

\begin{align}
e_{1}*\mathcal{K}_{\alpha}*\dots*\mathcal{K}_{\alpha}e_{-1} & =\frac{2\pi}{f'(\theta_{1})}\frac{(2\pi K_{\alpha}(0))^{k-2}}{f'(\theta_{1})^{k-2}}\Biggl(\frac{(2\pi K_{\alpha}(0))^{2}}{f'(\theta_{1})^{2}}+\frac{2\pi K_{\alpha}(0))}{f'(\theta_{1})}\times\\
 & \qquad i\int dm(v_{1})(K_{\alpha}(\theta_{1}-v_{1}+\frac{i\pi}{2})e^{\theta_{1}-v_{1}}-K_{\alpha}(v_{1}-\theta_{1}-\frac{i\pi}{2})e^{v_{1}-\theta_{1}})\nonumber \\
 & \qquad+(k-1)\frac{(2\pi)}{f'(\theta_{1})}\int dm(v_{1})K_{\alpha}(\theta_{1}-v_{1}+\frac{i\pi}{2})K_{\alpha}(v_{1}-\theta_{1}-\frac{i\pi}{2})\Biggr)\,,\nonumber 
\end{align}
where we kept only terms with at most one continuous convolution.
We need to sum the first line from $k=0$, the second from $k=1$
, while the last from $k=2$ to infinity. Also there is one more convolution
from (\ref{eq:1pLO}). Let us recall that 
\begin{equation}
f'(\theta_{1})=r\cosh\theta_{1}+2\pi K(0)-i\int dm(\theta)K(\theta_{1}+\frac{i\pi}{2}-\theta)\frac{\partial\epsilon(\theta)}{\partial\theta}
\,,\nn
\end{equation}
with the solution given by (\ref{eq:depstheta}). Expanding this formula
up to the leading exponential order and plugging back to the expressions
summed up agrees with (\ref{eq:1pNLO}). Let us emphasize that to obtain
the leading exponential contribution we need to sum up infinitely many terms in the discrete parts. Thus the agreement
found is a highly non-trivial test of our approach. 

\section{Small volume checks}

\begin{figure}
\begin{centering}
\includegraphics[width=3cm]{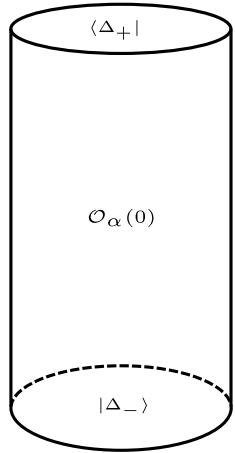}
\par\end{centering}
\caption{Cylindrical geometry for the conformal three-point functions.}

\label{cylinder}
\end{figure}

For small volume we compare the ratios of the expectation values to
the ratios of three-point functions in the Liouville conformal field
theory in a cylindrical geometry shown on Figure \ref{cylinder}.

The general three-point function in the CFT takes the form 
\begin{equation}
\langle\Delta_{+}\vert{\cal O}_{\alpha}(0)\vert\Delta_{-}\rangle=\langle\Delta\vert L_{n_{1}}\dots L_{n_{i}}(\mathbf{l}_{-m_{1}}\dots\mathbf{l}_{-m_{j}}\Phi_{\al})L_{-p_{1}}\dots L_{-p_{r}}\vert\Delta\rangle\,,\label{eq:cft3pt}
\end{equation}
where two different Virasoro modes are introduced. Both are related
to the same energy momentum tensor, but expanded around different
points.

Let us introduce a complex coordinate on the cylinder as $z=x+iy$
with $y\equiv y+2\pi$. By expanding $T(z)$ around the origin we
can act and change the operator, which is inserted. This action is
called the \emph{local} action: 
\begin{align}
T(z)=\sum_{n=-\infty}^{\infty}\mathbf{l}_{n}z^{-n-2} & \quad;\qquad\mathbf{l}_{n}\Phi_{\alpha}=\oint\frac{dz}{2\pi i}z^{n+1}T(z)\Phi_{\alpha}\,.\nn
\end{align}
For diagonal matrix elements, i.e. for expectation values, only even
mode numbers are used to generate the quotient space, where the action
of the conserved charges is factored out. 

By expanding $T(z)$ at $z\to\pm\infty$ we obtain the \emph{global
}action of the Virasoro algebra which can alter the initial and final
states: 
\begin{align}
T(z)=\sum_{n=-\infty}^{\infty}L_{n}e^{nz}-\frac{c}{24}\,.\nn
\end{align}
In order to relate the three-point function of the descendants (\ref{eq:cft3pt})
to that of the primary $\langle\Delta\vert \Phi_{\al}\vert\Delta\rangle\equiv\langle \Phi_{\al}\rangle_{\Delta}$
we use the cylinder conformal Ward identities:
\begin{align}
\langle T(z_{k})\cdots T(z_{1})\Phi_{\al}\rangle_{\Delta} & =-\frac{c}{12}\sum_{j=2}^{k}\chi'''(z_{1}-z_{j})\langle T(z_{k})\cdots\overset{j}{\widehat{\phantom{T}}}\cdots T(z_{2})\Phi_{\al}\rangle_{\Delta}\\
 & \quad+\Bigl\{\sum_{j=2}^{k}\bigl(-2\chi'(z_{1}-z_{j})+(\chi(z_{1}-z_{j})-\chi(z_{1}))\frac{\partial}{\partial z_{j}}\bigr)\nonumber \\
 & \quad\qquad\,\,\,\,\,\,\,\,\,-\Delta_{\al}\chi'(z_{1})+\Delta-\frac{c}{24}\Bigr\}\langle T(z_{k})\cdots T(z_{2})\Phi_{\al}\rangle_{\Delta}\nonumber 
\end{align}
where $\chi(z)=\frac{1}{2}\coth\left(\frac{z}{2}\right)\,$.

In calculating (\ref{eq:cft3pt}) we follow the prescription of \cite{Boos:2010ww}: we first take  $k=i+j+r$ and send
$z_{1},\dots,z_{i}$ to $-\infty$, $z_{i+1},\dots,z_{i+j}$ to $0$,
while $z_{i+j+1},\dots,z_{k}$ to $\infty$. By picking up the coefficient
of the appropriate power of $e^{\pm z}$ at $\mp\infty$ and $z$
around $0$ the three-point function (\ref{eq:cft3pt}) can be calculated.
In the following we first analyze non-degenerate $L_{0}$ subspaces,
i.e. highest weight states and their first descendants, and then level
$2$ states.

\subsection{Non-degenerate $L_{0}$ eigenspaces}

We perform this analysis for the low lying operators and states with
a non-degenerate $L_{0}$. This includes the state $\vert\Delta\rangle$
and $\vert\Delta+1\rangle\equiv L_{-1}\vert\Delta\rangle$, thus from
the table (\ref{tab}) we take all rows with $L=1,2,3,4$ and $M=0,1,\bar{M}=0$.
The computation using Ward identities provides 
\begin{align}
\frac{\langle\mathbf{l}_{-2}\Phi_{\al}\rangle_{\Delta}}{\langle \Phi_{\al}\rangle_{\Delta}} & =\Delta-\frac{c}{24}-\frac{\Delta_{\al}}{12}\,,\label{different}\\
\frac{\langle\mathbf{l}_{-4}\Phi_{\al}\rangle_{\Delta}}{\langle \Phi_{\al}\rangle_{\Delta}} & =\frac{\Delta_{\al}}{240}\,,\nonumber \\
\frac{\langle\mathbf{l}_{-2}^{2}\Phi_{\al}\rangle_{\Delta}}{\langle \Phi_{\al}\rangle_{\Delta}} & =\Delta^{2}-\Delta\frac{2\Delta_{\al}+c+2}{12}+\frac{20\Delta_{\al}^{2}+56\Delta_{\al}+20c\Delta_{\al}+5c^{2}+22c}{2880}\,.\nonumber 
\end{align}

\begin{align}
\frac{\langle \Phi_{\al}\rangle_{\Delta+1}}{\langle \Phi_{\al}\rangle_{\Delta}}\quad & =2\Delta+\Delta_{\al}^{2}-\Delta_{\al}\,,\label{eq:different2}\\
\frac{\langle\mathbf{l}_{-2}\Phi_{\al}\rangle_{\Delta+1}}{\langle \Phi_{\al}\rangle_{\Delta}} & =2\Delta^{2}+\Delta\frac{12\Delta_{\al}^{2}+34\Delta_{\al}+24-c}{12}-\frac{(\Delta_{\al}-1)\Delta_{\al}(2\Delta_{\al}-24+c)}{24}\,,\nonumber \\
\frac{\langle\mathbf{l}_{-4}\Phi_{\al}\rangle_{\Delta+1}}{\langle \Phi_{\al}\rangle_{\Delta}} & =\Delta\frac{241\Delta_{\al}}{120}-\frac{\Delta_{\al}^{2}}{240}+\frac{\Delta_{\al}^{3}}{240}\,,\nonumber \\
\frac{\langle\mathbf{l}_{-2}^{2}\Phi_{\al}\rangle_{\Delta+1}}{\langle \Phi_{\al}\rangle_{\Delta}} & =2\Delta^{3}+\Delta^{2}\frac{70-c+40\Delta_{\al}+6\Delta_{\al}^{2}}{6}\nonumber \\
 & \,\,\,+\Delta\frac{2400-218c+5c^{2}+4616\Delta_{\al}-340c\Delta_{\al}+1940\Delta_{\al}^{2}-120c\Delta_{\al}^{2}-240\Delta_{\al}^{3}}{1440}\nonumber \\
 & \,\,\,+\frac{(\Delta_{\al}-1)\Delta_{\al}(2400-218c+5c^{2}-424\Delta_{\al}+20c\Delta_{\al}+20\Delta_{\al}^{2})}{2880}\,\,.\nonumber 
\end{align}
where the central charge and the scaling dimensions of the operator and of the asymptotical state
are
\begin{equation}
c=1+6Q^{2},\quad\Delta_{\al}=\frac{Q^2} 4\al(2-\al),\quad \Delta=\frac{P^2} 2+\frac{Q^2} 4\,.
\end{equation}
We also need the ratio of the three-point functions for the primary
fields:
\begin{align}
\frac{\langle \Phi_{\al-2p}\rangle_{\Delta}}{\langle \Phi_{\al}\rangle_{\Delta}}=\frac{\gamma^{2}(ab-b^{2})}{\gamma(2ab-2b^{2})\gamma(2ab-b^{2})}\gamma(ab-b^{2}-2ibP)\gamma(ab-b^{2}+2ibP)\,;\quad a=\frac{\al Q} 2\,.
\end{align}
where $\gamma(x)=\Gamma(x)/\Gamma(1-x)$, and we used notations close to \cite{Zamolodchikov:1995aa} in order to simplify comparison.

Using the results of \cite{HGSIV,HGSV} we can relate the fermionic
basis to the low lying Virasoro descendants as 
\begin{align}
 & \Omega_{1,1}\simeq r^{-2}D_{1}(\al,p)D_{1}(2-\al,p)\frac{\langle\mathbf{l}_{-2}\Phi_{\al}\rangle}{\langle \Phi_{\al}\rangle}\,,\label{ch11}\\
 & \Omega_{3,1}\simeq r^{-4}\frac{1}{2}D_{3}(\al,p)D_{1}(2-\al,p)\Bigl\{\frac{\langle\mathbf{l}_{-2}^{2}\Phi_{\al}\rangle}{\langle \Phi_{\al}\rangle}+\Bigl(\frac{2c-32}{9}+\frac{2}{3}d(\al,p)\Bigr)\frac{\langle\mathbf{l}_{-4}\Phi_{\al}\rangle}{\langle \Phi_{\al}\rangle}\Bigr\}\,,\label{ch13}\\
 & \Omega_{1,3}\simeq r^{-4}\frac{1}{2}D_{1}(\al,p)D_{3}(2-\al,p)\Bigl\{\frac{\langle\mathbf{l}_{-2}^{2}\Phi_{\al}\rangle}{\langle \Phi_{\al}\rangle}+\Bigl(\frac{2c-32}{9}-\frac{2}{3}d(\al,p)\Bigr)\frac{\langle\mathbf{l}_{-4}\Phi_{\al}\rangle}{\langle \Phi_{\al}\rangle}\Bigr\}\,,\label{ch31}\\
 & \Omega_{1,-1}\simeq r^{2(\Delta_{\al}-\Delta_{\al-b})}t_{1}(\al)F(\al,p)\frac{\langle \Phi_{\al-2p}\rangle}{\langle \Phi_{\al}\rangle}\,,\label{chprim}
\end{align}
where 
$$d(\al,p)=\frac{2p-1}{p(p-1)}(\al-1)\,,$$
and the expectation values are taken
in the finite volume eigenstate of the conserved charges. The appearing
coefficients for descendants originate from the normalization of the
fermionic operators 
\begin{align}
D_{m}(\al,p)
=\frac{1}{2i\sqrt{\pi}}Z(p)^{-m}\Gamma\Bigl(\frac {\al+mp} 2\Bigr)\Gamma\Bigl(\frac {\al+m(1-p)} 2\Bigr)\,,
\end{align}
while for primaries they are essentially the ratio of two Lukyanov-Zamolodchikov
one-point functions
\begin{align}
F(\al,p)=Z(p)^{2(\Delta_{\al}-\Delta_{\al-2p})}\frac 2 {1-p}\ \gamma\Bigl(\frac{\al+1-p}{2}\Bigr)\gamma\Bigl(\frac{2-\al+p}{2}\Bigr)\gamma\Bigl(\frac{\al-p}{1-p}\Bigr)\,,
\end{align}

For asymptotical states we consider either primary fields parametrized
by the quantum number $L$ or their first descendants. For the primary
fields the formulae above are taken literally. For the descendants
we have to use the formulae (\ref{different}) carefully as, for instance,
\begin{align}
\frac{\langle\mathbf{l}_{-2}\Phi_{\al}\rangle_{\Delta+1}}{\langle \Phi_{\al}\rangle_{\Delta+1}} & =\frac{\langle\mathbf{l}_{-2}\Phi_{\al}\rangle_{\Delta+1}}{\langle \Phi_{\al}\rangle_{\Delta}}\frac{\langle \Phi_{\al}\rangle_{\Delta}}{\langle \Phi_{\al}\rangle_{\Delta+1}}\nonumber \\
 & =\frac{48\Delta^{2}+2\Delta(12\Delta_{\al}^{2}+34\Delta_{\al}+24-c)-{(\Delta_{\al}-1)\Delta_{\al}(2\Delta_{\al}-24+c)}}{{24}(2\Delta+\Delta_{\al}^{2}-\Delta_{\al})}\,.
\end{align}

In the Table 1 we compare the numerical values of $\Omega_{i,j}$ obtained
for 
\begin{align}
r=.001,\ \ a=\frac{87}{80},\ \ b=\frac{2}{5}\,,\label{don}
\end{align}
to their CFT limits.

\begin{table}[h]
\begin{centering}
\begin{tabular}{|c|c|c|c|c|}
\hline 
state & \multicolumn{1}{c}{$M=0$} & $L=1$ & \multicolumn{1}{c}{$M=1$} & $L=1$\tabularnewline
\hline 
 & numerical & CFT & \multicolumn{1}{c}{numerical} & CFT\tabularnewline
\hline 
\hline 
$\Omega_{1,1}$ & $\ensuremath{3.85677\cdot10^{6}}$ & $3.85677\cdot10^{6}$ & $-6.60202\cdot10^{7}$ & $\ensuremath{-6.60203\cdot10^{7}}$\tabularnewline
\hline 
$\Omega_{3,1}$ & $\ensuremath{1.00405\cdot10^{14}}$ & $1.00405\cdot10^{14}$ & $1.07476\cdot10^{16}$ & $1.07475\cdot10^{16}$\tabularnewline
\hline 
$\Omega_{1,3}$ & $1.04361\cdot10^{14}$ & $1.04361\cdot10^{14}$ & $1.05988\cdot10^{16}$ & $1.05987\cdot10^{16}$\tabularnewline
\hline 
$\Omega_{1,-1}$ & $-0.0028363$ & $-0.0028363$ & $-0.00231607$ & $-0.00231668$\tabularnewline
\hline 
\end{tabular}
\par\end{centering}
\vspace{.5cm}
\begin{centering}
\begin{tabular}{|c|c|c|c|c|}
\hline 
state & \multicolumn{1}{c}{$M=0$} & $L=2$ & \multicolumn{1}{c}{$M=1$} & $L=2$\tabularnewline
\hline 
 & numerical & CFT & \multicolumn{1}{c}{numerical} & CFT\tabularnewline
\hline 
\hline 
$\Omega_{1,1}$ & $\ensuremath{3.79053\cdot10^{6}}$ & $3.79053\cdot10^{6}$ & $-6.61159\cdot10^{7}$ & $\ensuremath{-6.61132\cdot10^{7}}$\tabularnewline
\hline 
$\Omega_{3,1}$ & $\ensuremath{9.93725\cdot10^{13}}$ & $9.93725\cdot10^{13}$ & $1.0771\cdot10^{16}$ & $1.07703\cdot10^{16}$\tabularnewline
\hline 
$\Omega_{1,3}$ & $1.03188\cdot10^{14}$ & $1.03188\cdot10^{14}$ & $1.06245\cdot10^{16}$ & $1.06237\cdot10^{16}$\tabularnewline
\hline 
$\Omega_{1,-1}$ & $-0.00276414$ & $-0.00276451$ & $-0.00225609$ & $-0.00225862$\tabularnewline
\hline 
\end{tabular}
\par\end{centering}
\vspace{.5cm}
\begin{centering}
\begin{tabular}{|c|c|c|c|c|}
\hline 
state & \multicolumn{1}{c}{$M=0$} & $L=3$ & \multicolumn{1}{c}{$M=1$} & $L=3$\tabularnewline
\hline 
 & numerical & CFT & \multicolumn{1}{c}{numerical} & CFT\tabularnewline
\hline 
\hline 
$\Omega_{1,1}$ & $3.68203\cdot10^{6}$ & $3.68197\cdot10^{6}$ & $-6.62725\cdot10^{7}$ & $\ensuremath{-6.62653\cdot10^{7}}$\tabularnewline
\hline 
$\Omega_{3,1}$ & $\ensuremath{9.77084\cdot10^{13}}$ & $9.77074\cdot10^{13}$ & $1.08094\cdot10^{16}$ & $1.08076\cdot10^{16}$\tabularnewline
\hline 
$\Omega_{1,3}$ & $1.01297\cdot10^{14}$ & $1.01296\cdot10^{14}$ & $1.06668\cdot10^{16}$ & $1.06648\cdot10^{16}$\tabularnewline
\hline 
$\Omega_{1,-1}$ & $-0.00265524$ & $-0.00265529$ & $-0.00216529$ & $-0.0021703$\tabularnewline
\hline 
\end{tabular}
\par\end{centering}
\vspace{.5cm}
\begin{centering}
\begin{tabular}{|c|c|c|c|c|}
\hline 
state & \multicolumn{1}{c}{$M=0$} & $L=4$ & \multicolumn{1}{c}{$M=1$} & $L=4$\tabularnewline
\hline 
 & numerical & CFT & \multicolumn{1}{c}{numerical} & CFT\tabularnewline
\hline 
\hline 
$\Omega_{1,1}$ & $3.53306\cdot10^{6}$ & $3.53306\cdot10^{6}$ & $-6.64869\cdot10^{7}$ & $-6.64737\cdot10^{7}$\tabularnewline
\hline 
$\Omega_{3,1}$ & $9.54791\cdot10^{13}$ & $9.5479\cdot10^{13}$ & $1.08622\cdot10^{16}$ & $1.08589\cdot10^{16}$\tabularnewline
\hline 
$\Omega_{1,3}$ & $9.87646\cdot10^{13}$ & $9.87645\cdot10^{14}$ & $1.07248\cdot10^{16}$ & $1.07212\cdot10^{16}$\tabularnewline
\hline 
$\Omega_{1,-1}$ & $-0.00251988$ & $-0.00252043$ & $-0.00205411$ & $-0.00206125$\tabularnewline
\hline 
\end{tabular}
\par\end{centering}
\caption{We calculate numerically $\Omega_{i,j}$ for various states and compare
them to their exact conformal counterparts. }
\end{table}

\subsection{Checks with degenerate $L_{0}$ spaces}\label{degenerate}

Here we consider the simplest case of degeneracy: level 2. We work
in the basis: 
\begin{equation}
L_{-2}|\Delta\rangle,\quad L_{-1}^{2}|\Delta\rangle\,.\label{eq:level2basis}
\end{equation}
There are two eigenvectors of the local integrals of motion. Since
$I_{1}=L_{0}-c/24$ does not distinguish between them we consider
the next conserved charge: 
\begin{equation}
I_{3}=2\sum\limits _{n=1}^{\infty}L_{-n}L_{n}+L_{0}^{2}-\frac{c+2}{12}L_{0}+\frac{c(5c+22)}{2880}\,.
\end{equation}
This integral of motion is a $2\times2$ matrix in the basis above,
with eigenvalues 
\begin{align}
\lambda_{\pm}(\Delta)=\frac{17}{3}+\frac{c(5c+982)}{2880}-\frac{c-142}{12}\Delta+\Delta^{2}\pm\frac{1}{2}\sqrt{288\Delta+(c-4)^{2}}\,,
\end{align}
and eigenvectors 
\begin{equation}
\psi_{\pm}=\begin{pmatrix}\frac{1}{12}\ensuremath{c-4\pm\sqrt{288\Delta+(c-4)^{2}}}\\
1
\end{pmatrix}
\end{equation}
For simplicity we consider $L=1$. In table (\ref{tab}) we present
two cases with $L=1,M=2,\bar{M}=0$: $\{1,3\}$, $\{4\}$. We first
identify which one corresponds to $\lambda_{+}$ and which one to
$\lambda_{-}$. In doing so we recall the general eigenvalue of the
local integral of motion: 
\begin{align}
I_{n}(r)=\frac{1}{C_{n}(p)}\Bigl(-\frac{1}{n}\sum_{j=1}^{m}e^{n\theta_{k}}+(-1)^{\frac{n-1}{2}}\frac{1}{2\pi}\int\limits _{-\infty}^{\infty}e^{n\theta}\log\Bigl(1+e^{-\epsilon(\theta)}\Bigr)d\theta\Bigr)\,,
\end{align}
where 
\begin{equation}
C_{n}(p)=-\frac{Z(p)^{-n}}{4\sqrt{\pi}Q\ensuremath{\frac{n+1}{2}}!}\Gamma(np)\Gamma(n(1-p))\,.
\end{equation}
The normalized eigenvalue $\tilde{I}_{n}(r)=r^{n}I_{n}(R)$ should
approach the CFT limit. For $r=10^{-3}$ we obtained the following
numerical results:
\begin{align}
 & \mathcal{N}_{-}=\{1,3\}\,,\quad\tilde{I}_{3}(r)=21.3773\,,\quad\lambda_{-}(\Delta(r))=21.3767\,,\\
 & \mathcal{N}_{+}=\{4\}\,,\ \ \ \quad\tilde{I}_{3}(r)=74.8405\,,\quad\lambda_{+}(\Delta(r))=74.8399\,.\nonumber 
\end{align}
which establishes the required correspondence.

For any local operator \emph{$\mathcal{O}$} we introduce a $2\times2$
matrix, which contains its matrix elements in the basis (\ref{eq:level2basis}).
We denote this matrix by $\frac{\langle\mathbf{l}_{-2}O\rangle_{\Delta+2}}{\langle \Phi_{\al}\rangle_{\Delta}}$.
We need the following two cases

\begin{align}
\frac{\langle \Phi_{\al}\rangle_{\Delta+2}}{\langle \Phi_{\al}\rangle_{\Delta}} & =\begin{pmatrix}4\Delta-4\Delta_{\al}+4\Delta_{\al}^{2}+\frac{c}{2} & 2(3\Delta-\Delta_{\al}+\Delta_{\al}^{3})\\
2(3\Delta-\Delta_{\al}+\Delta_{\al}^{3}) & 8\Delta^{2}+\Delta(4-8\Delta_{\al}+8\Delta_{\al}^{2})-2\Delta_{\al}+3\Delta_{\al}^{2}-2\Delta_{\al}^{3}+\Delta_{\al}^{4}
\end{pmatrix}\,.
\end{align}
and
\begin{equation}
\frac{\langle\mathbf{l}_{-2}\Phi_{\al}\rangle_{\Delta+2}}{\langle \Phi_{\al}\rangle_{\Delta}}=\begin{pmatrix}M_{1,1} & M_{1,2}\\
M_{1,2} & M_{2,2}
\end{pmatrix}\nn\,,
\end{equation}
with entries
\begin{align}
M_{1,1} & =\frac{1}{48}\bigl(48c-c^{2}-672\Delta_{\al}+102c\Delta_{\al}+976\Delta_{\al}^{2}-8c\Delta_{\al}^{2}-16\Delta_{\al}^{3}\nonumber \\
 & \quad\quad+\Delta(384+16c+560\Delta_{\al}+192\Delta_{\al}^{2})+192\Delta^{2}\bigr)\,,\nonumber \\
M_{1,2} & =\frac{1}{12}\bigl(-72\Delta_{\al}+7c\Delta_{\al}+14\Delta_{\al}^{2}+6c\Delta_{\al}^{2}+84\Delta_{\al}^{3}-c\Delta_{\al}^{3}-2\Delta_{\al}^{4}\nonumber \\
 & \quad\quad+\Delta(144-3c+258\Delta_{\al}+144\Delta_{\al}^{2}+24\Delta_{\al}^{3})+72\Delta^{2}\bigr)\,,\nonumber \\
M_{2,2} & =\frac{1}{24}\bigl(-96\Delta_{\al}+2c\Delta_{\al}+52\Delta_{\al}^{2}-3c\Delta_{\al}^{2}-6\Delta_{\al}^{3}+2c\Delta_{\al}^{3}+52\Delta_{\al}^{4}-c\Delta_{\al}^{4}-2\Delta_{\al}^{5}\nonumber \\
 & \quad\quad+\Delta(192-4c+232\Delta_{\al}+8c\Delta_{\al}+568\Delta_{\al}^{2}-8c\Delta_{\al}^{2}+128\Delta_{\al}^{3}+24\Delta_{\al}^{4})\nonumber \\
 & \quad\quad+\Delta^{2}(480-8c+560\Delta_{\al}+192\Delta_{\al}^{2})+192\Delta^{3}\bigr)\,.\nn
\end{align}

We now rewrite the general formulae (\ref{ch11}), (\ref{chprim})
for the present case 
\begin{align}
 & \Omega_{1,1}^{\pm}\simeq r^{-2}D_{1}(\al,p)D_{1}(2-\al,p)\frac{\psi_{\pm}^{t}\cdot\langle\mathbf{l}_{-2}\Phi_{\al}\rangle_{\Delta+2}\cdot\psi_{\pm}}{\psi_{\pm}^{t}\cdot\langle \Phi_{\al}\rangle_{\Delta+2}\cdot\psi_{\pm}}\,,\nonumber \\
 & \Omega_{1,-1}^{\pm}\simeq r^{2(\Delta_{\al}-\Delta_{\al-2p})}t_{1}(a,b)F(\al,p)\frac{\psi_{\pm}^{t}\cdot\langle \Phi_{\al-2p}\rangle_{\Delta+2}\cdot\psi_{\pm}}{\psi_{\pm}^{t}\cdot\langle \Phi_{\al}\rangle_{\Delta+2}\cdot\psi_{\pm}}\,.\nn
\end{align}
We compute these quantities at the numerical values (\ref{don}).
The results are summarized in the table 
\begin{center}
\begin{tabular}{|c||c|c||c|c|}
\hline 
eigenvalue & $\Omega^{-}$ & CFT & $\Omega^{+}$ & CFT\tabularnewline
\hline 
\hline 
$\Omega_{1,1}$ & $-1.08278\cdot10^{8}$ & $-1.08276\cdot10^{8}$ & $\ensuremath{-1.88722\cdot10^{8}}$ & $-1.88716\cdot10^{8}$\tabularnewline
\hline 
$\Omega_{1,-1}$ & $\ensuremath{-0.00210992}$ & $-0.00211103$ & $-0.00245252$ & $-0.00245289$\tabularnewline
\hline 
\end{tabular}
\par\end{center}

Thus we see that our procedure works well in the case with degenerate
$L_{0}$, too. This completes the small volume check of our conjecture. 

\section{Conclusions}

We conjectured compact expressions for the finite volume diagonal
matrix elements of exponential operators and their descendants in
the sinh-Gordon theory. By using the fermionic basis to create the
descendant operators we could relate their finite volume expectation
values to that of the primaries in terms of a determinant with entries,
which satisfies a linear integral equation. Careful choice of the
fermionic creation operators can relate the matrix elements of two
different exponential operators allowing, in principle, they complete
determination. The linear integral equation contains a measure, which
is built up from the pseudo-energy of the excited state TBA equations
and a kernel, which is a deformation of the TBA kernel. Excited states
are characterized by the discrete rapidities of the particles and
the continuous pseudo-energy and the two parts are connected by the
TBA and BA equations. They both appear in the linear integral equations,
which can be solved by iterations. The discrete part is responsible
for the polynomial finite size corrections, while the continuous part
for the exponentially small ones. We checked for low number of particles
that summing up all the polynomial corrections the asymptotic diagonal
finite volume form factors can be recovered. We also checked the leading
exponential correction against Pozsgay's formula and found complete
agreement. The integral equation can also be solved numerically. The
small volume limit of the solution allows us to map multi-particle
states to the spectrum of the Liouville conformal field theory and
compare our conjecture to the CFT three-point functions providing
ample evidence for its correctness.

In calculating the asymptotic expressions for the finite volume form
factors we used a deformation of the TBA kernel, which is the logarithmic
derivative of the scattering matrix. We believe that this alternative
form for the connected and symmetric form factors can be used to find
a compact and closed expression for them and we initiate a study into
this direction.

It would be very nice to extend our exact finite volume results for
non-diagonal form factors. These results then could be tested for large
volumes against the leading exponential correction of form factors 
\cite{Bajnok2019prep}.

Finally, the knowledge of all form factors could give rise
to the determination of finite volume correlation functions relevant both in statistical
and solid state physics.

It is an interesting question whether the very nice structure we obtained
for the sinh-Gordon model extends to other integrable models such
as O(N) models or the AdS/CFT correspondence.

\subsection*{Acknowledgments}

This research was supported by the NKFIH research Grant K116505, by the
Lend\"ulet Program of the Hungarian Academy of Sciences and by a Hungarian-French
bilateral exchange project. FS is grateful to Wigner Research Centre
for Physics where this work was started for kind hospitality.

\end{document}